# Simulation Study  For Performance  Comparison in Hierarchical Network With CHG Approach in MANET

Anzar ahamd[1,1], Prof R. Gowri[1], Prof. S.C.Gupta [2],

[1] Deptt of  Electronics & Communication,
Graphic Era University,
Dehradun

[2] Deptt of  Computer Science,
IIT Roorkee ,
Roorkee

Anz.hmd@gmail.com
R.Gowri@gmail.com
sureshprem1938@gmai.com

**Abstract.** The implementation of MANET for commercial purposes is not an easy task. Unlike other wireless technologies such as cellular networks, MANET face more difficult problems concerning management functions, routing and scalability . As a solution to these complications, clustering schemes are proposed for MANET in order to organize the network topology in a hierarchical manner. Many clustering techniques have been developed. Clustering is a method which aggregates nodes into groups . These groups are contained by the network and they are known as clusters.   By Increasing network capacity and reducing the routing overhead through clustering brings more efficiency and effectiveness to scalability in relation to node numbers and the necessity for high mobility. The manager node,in clustering has responsibility for many functions such as cluster maintenance, routing table updates, and the discovery of new routes within the network.    The other node   named as gateway node communicate to the other cluster . In this  paper we remove the cluster head  (CH)   and given a new approach in which cluster head and gateway will be same and that node is known as cluster head gateway (CHG) ,in which all the responsibilities of cluster head and gateway will be perform by the  Cluster head gateway(CHG) itself . By applying this approach we  reduce of overheads and  improve the over all performance  of the network while throughput will be same in both condition with the help of Exata simulation.

---

[2] Anzar Ahamd.

**Keywords:** MANET, CH,CHG, Overheads,

# 1  Introduction

Now a days, wireless technologies are becoming quite common in our daily lives. They have been gaining popularity with the use of portable devices like laptop computers, personal digital assistants, and mobile phones. In order to use these devices some type of fixed infrastructure is normally required such as access points or base stations. This means that unless mobile users of wireless technologies have the possibility to access a static network, they will not be able to support their mobile devices services. Mobile ad hoc networks  (MANET) propose a solution to these kinds of problems. MANET are autonomous systems consisting of mobile hosts that are connected by multi-hop wireless links[29]. The main idea of a MANET is that a network can be established without the need for any centralized administration or fixed infrastructure. MANET present many challenges to the research community because of dynamic topologies. In addition, link bandwidth and mobile nodes transmission power are scarce . Also Scalability is of particular interest to ad hoc network designers and users and is an issue with critical influence on capability and capacity. The scalability issue of MANET is addressed through a hierarchical approach that partition the network into clusters. A cluster is basically a subset of nodes of the network that satisfies a certain property . Clusters are analogous to cells in a cellular network. In this way the network becomes more  manageable[29][30].  . It must be clear though that a clustering technique is not a routing protocol. Clustering is a method which aggregates nodes into groups [4]. These groups are contained by the network and they are known as clusters. A cluster is basically a subset of nodes of the network that satisfies a certain property[1]. Each cluster has some nodes and  a cluster leader or head which look after other nodes in the cluster and a gateway node which communicate to other cluster .In this way the network becomes more manageable[25][26][27][28] .

## 2  Related Work

Routing is one of the major challenges in MANET. Routing in MANET has three major goals. [7] ,one gives the maximum reliability by selecting alternatives route if a node fails, second ,route  traffic through the path with least cost in the network and third give the nodes the best possible response time and throughput. This is specially important for the interactive session between user application.Routing can be classified in MANET as proactive and reactive. In proactive routing routers attempts continuously the routes within the network. In reactive protocol invoke the route determination procedure only on demand. Cluster based routing [9] is a convenient way for routing in MANET . In MANET nodes are very close to each other normally one hop or two hop distance, each cluster has one or more gateway node to connect to other cluster in the network[1].Back bone . baserouting[10]andspinebased routing [11],[1] uses a similar approach. Clustering presents several advantages for the medium access layer and the network layer in MANET [21]. The implementation of clustering schemes allow a better performance of the protocols for the Medium Access Control (MAC) layer by improving the spatial reuse, throughput, scalability and power consumption. On the other hand, clustering helps improve routing at the network layer by reducing the size of the routing tables and by decreasing transmission overhead due to the update of routing tables after topological changes occur [1][23][20]. Clustering helps aggregate topology information since the number of nodes of a cluster is smaller than the number of nodes of the entire network[9][13][14]. Therefore, each node only needs to store a fraction of the total network routing information [1][22]. In most clustering techniques nodes are selected to play different roles according to a certain criteria. In general, three types of nodes are defined: Ordinary nodes are simply  members of a cluster ,other node is  are called gateways node because they are able to listen to transmissions from another node which is in a different cluster[1] [22][27][30]. In most clustering techniques nodes are selected to play different roles according to a certain criteria. In general, three types of nodes are defined:

## 2.1 Ordinary nodes

Ordinary nodes are members of a cluster which do not have neighbours belonging to a different cluster [1].

## 2.2 Gateway nodes

Gateway nodes are nodes in a non-clusterhead state located at the periphery of a cluster. These types of nodes are called gateways because they are able to listen to transmissions from

## 2.3  Clusterheads

Most clustering approaches for mobile ad hoc networks select a subset of nodes in order to form a network backbone that supports control functions. A set of the selected nodes are called clusterhead(CH) and each node in the network is associated with one. Clusterheads are connected with one another directly or through gateway nodes[1]. The union of gateway nodes and clusterheads form a connected backbone. This connected backbone helps simplify functions such as channel access,bandwidth allocation, routing power control and virtual-circuit support [18].another node which is in a different cluster [21]. To accomplish this, a gateway node must have at least one neighbour that is a member of another cluster [24].

## 3  PROBLEM FORMULATION

In our approach we compared   cluster head(CH) and gateway enable architecture
which is shown in fig 1 and  cluster head gateway(CHG)  enable architecture which
will act as a cluster head as well as   gateway (CHG) as shown in the figure2  .

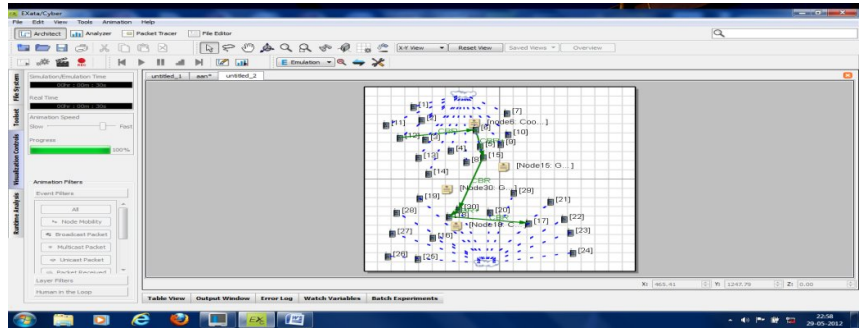

Fig1: CH& gateway  scenario

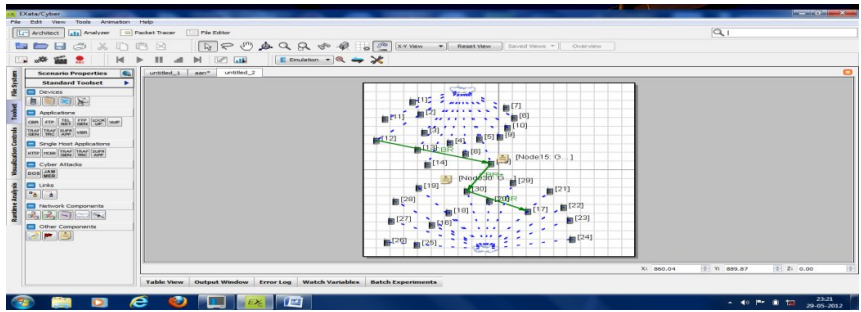

## 4   SIMULATION ENVIRONMENT

We used  Exata Cyber simulator   to simulate   CHG approach. EXata is a
comprehensive suite of tools for emulating large wired and wireless networks. It uses
simulation and emulation to predict the behavior and performance of networks to
improve their design, operation, and management. EXata SVN provides a cost-
effective and easy-to-use alternative to physical testbeds that typically have high
equipment costs, complex setup requirements and limited scalability.It creates a
digital network replica that interfaces with real networks and applications.

## 4.1  Problem Definition

We have taken 30 mobile nodes with AODV  enabled, and all nodes are  randomly
distributed with the mobility of 0- 10 me / sec. In both scenario Node 12 is source
node and node17 is destination node sending CBR file. Node  6 and 15  is Cluster
Heads  node and node 18,30 are gateway node in two cluster network  respectively .In
other  scenario  node  no.6  and  15  is replaced    with  CHG  node  no.  15  and
30.Environment size for simulation  is 1500x1500 mtrs .

## 4.2  Simulation Setup:

In  Exata first we have to configure the profile for MANET, and there were three
important configurations for standard application , mobility and placement, network
layer and Routing algorithm.

## 4.3   Mobility Configuration

Mobility configuration, related to description about the mobility of mobile no des, and
for this we set the three important parameters

TABLE 1
 MOBILITY CONFIGURATION

| S. No | Parameter | Value |
|---|---|---|
| 1 | Speed | 0-10 Meter / Sec |
| 2 | Pause Time | 0 Sec |
| 3 | Start Time | 10 Sec. |

## 5    VALIDATION

### 5.1 Analysis for  end-to end  delay in CH & G  and CHG Approach

In the result analysis of CBR server  with cluster head & gateway  , and  Cluster head gateway  (CHG)  the  end  to  end  delay  for  the  CHG  is  a  trade  off   for   overall performance of the  network and  as shown in fig 3 and fig4

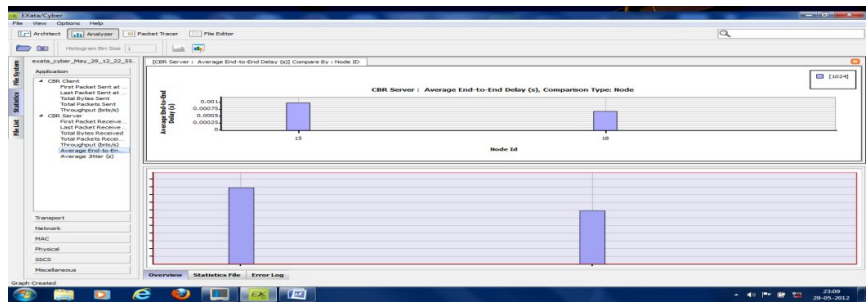

Fig3: Avg end to end delay for CH& G

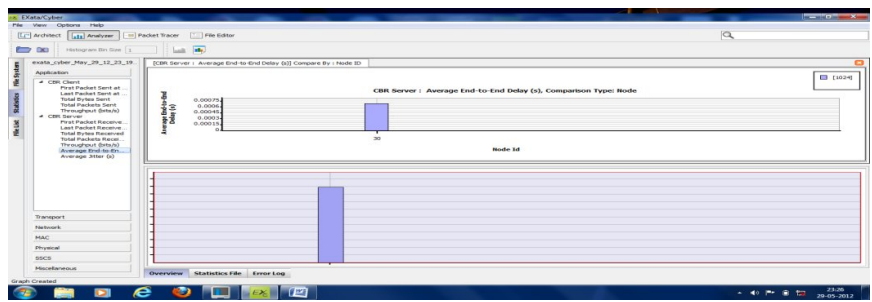

Fig4: Avg end to end delay in CH G

### 5.2  Analysis for Average Jitter with CH & G  and CHG Approach

In the result analysis of CBR server  with cluster head & gateway  , and  Cluster head gateway  (CHG)  the    average jitter is less in   CHG  approach  showing  over  all performance  of the node as shown in the  fig5 and fig6   and also less node is participating hence less overheads.

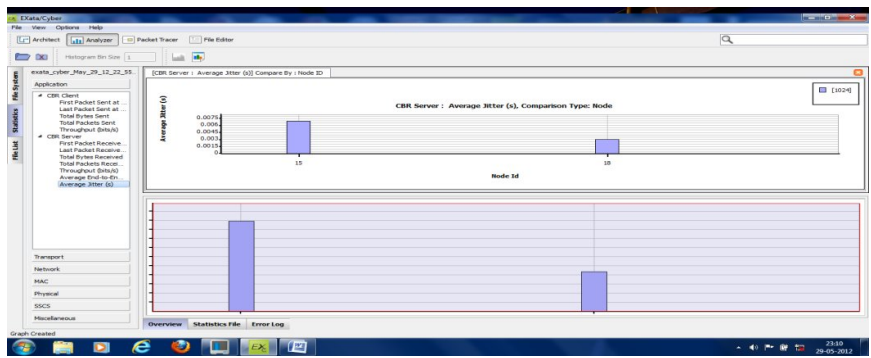

Fig5: Average  jitter  in CH& G

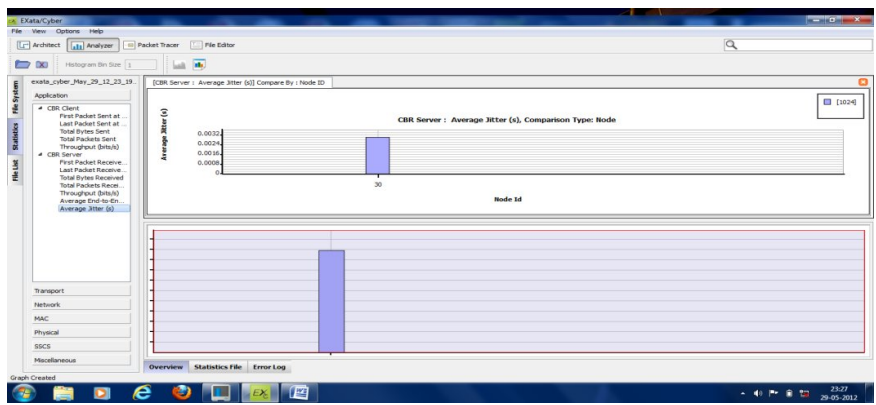

Fig6: Average  jitter  in CHG

### 5.3 Analysis for number of packets dropped in CH & G and CHG Approach

In the result analysis of CBR server with cluster head & gateway , and Cluster head gateway (CHG) the number of packets dropped in CHG approach is less and over all performance of the network is very much improved as shown in the figure

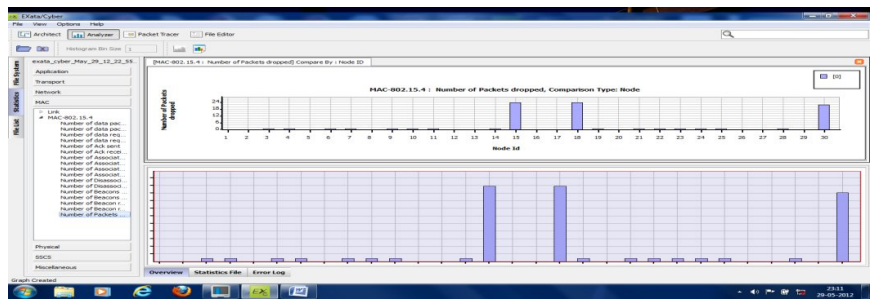

Fig7 :Mac 802.11.4 no. of pkt dropped in CH& G

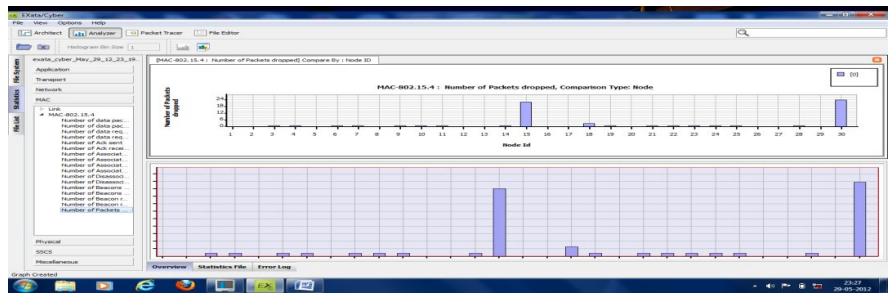

Fig8 :Mac 802.11.4 no. of pkt dropped in CH G

### 6 CONCLUSION

Wireless technology is gaining ground because of its low cost and ease-of-use Mobile Adhoc Network is next generation technology .In the given approach we have reduce the election of cluster head and election of suitable gateway with the CHG approach. Obviously the CHG terminal has to perform extra work but no. of dropped pakests and overheads are less at MAC layer.Hence by CGH approach overall performance may be increased .